\documentstyle[psfig]{mn}

\newif\ifAMStwofonts

\def\heII{He\,{\sc ii}\,$\lambda$4686}
\def\mgII{Mg\,{\sc ii}\,$\lambda$4481}
\def\hel2{He\,{\sc ii}}
\def\mg2{Mg\,{\sc ii}}
\def\he1{He\,{\sc i}}
\def\naI{Na\,{\sc i}\,$\lambda\lambda$8183/94}
\def\na1{Na\,{\sc i}}
\def\kmps{km\,s$^{-1}$}
\def\qq{QQ~Vul}

\def\fisp{\phi_{\rm spec}}

\ifoldfss
  \ifCUPmtlplainloaded \else
    \NewTextAlphabet{textbfit} {cmbxti10} {}
    \NewTextAlphabet{textbfss} {cmssbx10} {}
    \NewMathAlphabet{mathbfit} {cmbxti10} {} 
    \NewMathAlphabet{mathbfss} {cmssbx10} {} 
  \fi
  \ifAMStwofonts
    \ifCUPmtlplainloaded \else
      \NewSymbolFont{upmath} {eurm10}
      \NewSymbolFont{AMSa} {msam10}
      \NewMathSymbol{\upi}     {0}{upmath}{19}
      \NewMathSymbol{\umu}     {0}{upmath}{16}
      \NewMathSymbol{\upartial}{0}{upmath}{40}
      \NewMathSymbol{\leqslant}{3}{AMSa}{36}
      \NewMathSymbol{\geqslant}{3}{AMSa}{3E}

    \fi
  \fi
\fi 

\ifnfssone
  \newmathalphabet{\mathit}
  \addtoversion{normal}{\mathit}{cmr}{m}{it}
  \addtoversion{bold}{\mathit}{cmr}{bx}{it}
  \newmathalphabet{\mathbfit} 
  \addtoversion{normal}{\mathbfit}{cmr}{bx}{it}
  \addtoversion{bold}{\mathbfit}{cmr}{bx}{it}
  \newmathalphabet{\mathbfss} 
  \addtoversion{normal}{\mathbfss}{cmss}{bx}{n}
  \addtoversion{bold}{\mathbfss}{cmss}{bx}{n}
  \ifAMStwofonts
    \ifCUPmtlplainloaded \else
      \UseAMStwoboldmath
      \makeatletter
      \new@mathgroup\upmath@group
      \define@mathgroup\mv@normal\upmath@group{eur}{m}{n}
      \define@mathgroup\mv@bold\upmath@group{eur}{b}{n}
      \edef\UPM{\hexnumber\upmath@group}
      \new@mathgroup\amsa@group
      \define@mathgroup\mv@normal\amsa@group{msa}{m}{n}
      \define@mathgroup\mv@bold\amsa@group{msa}{m}{n}
      \edef\AMSa{\hexnumber\amsa@group}
      \makeatother
      \mathchardef\upi="0\UPM19
      \mathchardef\umu="0\UPM16
      \mathchardef\upartial="0\UPM40
      \mathchardef\leqslant="3\AMSa36
      \mathchardef\geqslant="3\AMSa3E
    \fi
  \fi
\fi 

\ifnfsstwo
  \DeclareMathAlphabet{\mathbfit}{OT1}{cmr}{bx}{it}
  \SetMathAlphabet\mathbfit{bold}{OT1}{cmr}{bx}{it}
  \DeclareMathAlphabet{\mathbfss}{OT1}{cmss}{bx}{n}
  \SetMathAlphabet\mathbfss{bold}{OT1}{cmss}{bx}{n}
  \ifAMStwofonts
    \ifCUPmtlplainloaded \else
      \DeclareSymbolFont{UPM}{U}{eur}{m}{n}
      \SetSymbolFont{UPM}{bold}{U}{eur}{b}{n}
      \DeclareSymbolFont{AMSa}{U}{msa}{m}{n}
      \DeclareMathSymbol{\upi}{0}{UPM}{"19}
      \DeclareMathSymbol{\umu}{0}{UPM}{"16}
      \DeclareMathSymbol{\upartial}{0}{UPM}{"40}
      \DeclareMathSymbol{\leqslant}{3}{AMSa}{"36}
      \DeclareMathSymbol{\geqslant}{3}{AMSa}{"3E}
    \fi
  \fi
\fi 

\ifCUPmtlplainloaded \else
  \ifAMStwofonts \else 
    \def\upi{\pi}
    \def\umu{\mu}
    \def\upartial{\partial}
  \fi
\fi

\title[Multi-epoch Doppler tomography and polarimetry of QQ Vul]
{Multi-epoch Doppler tomography and polarimetry of QQ Vul
\thanks{Based in part on observations at the
European Southern Observatory La Silla (Chile) with the 2.2m telescope 
of the Max-Planck-Society}}

\author[A. D. Schwope et al.]
	{Axel D.~Schwope$^1$\thanks{Visiting 
		astronomer,
              German-Spanish Astro\-no\-mi\-cal Center, Calar Alto, 
              operated by the Max-Planck-Institut f\"ur Astronomie, 
              Heidelberg, jointly with the Spanish National Commission 
              for Astronomy.},
           Maria S. Catal\'an$^{2}$,
           Klaus Beuermann$^{3,4}$,
	   Andr\'e Metzner$^{1}$,
	\newauthor
	   Robert Connon Smith$^{5}$,
	   Danny Steeghs$^{6}$\\
	$^1$Astrophysikalisches Institut Potsdam, An der Sternwarte 16,
        14482 Potsdam, Germany\\
	$^2$Keele University, Department of Physics, Keele, Staffordshire, 
	ST5 5BG, UK\\
	$^3$Universit\"{a}tssternwarte G\"{o}ttingen, Geismarlandstrasse 11,
        D--37083 G\"{o}ttingen, Germany\\ 
        $^4$Max-Planck-Institut f\"{u}r Extraterrestrische Physik, 
        Karl-Schwarzschild-Strasse 1, D-85748 Garching, Germany \\
	$^5$Astronomy Centre, School of CPES,
	University of Sussex, Falmer,  Brighton BN1 9QJ, UK\\
	$^6$Univ.~of St.~Andrews, School of Physics and Astronomy, 
        North Haugh, St.~Andrews, Fife KY16 9SS, Scotland, UK}

\date{Accepted.
      Received 1999 March;
      in original form 1999}

\pagerange{\pageref{firstpage}--\pageref{lastpage}}
\pubyear{1999}

\begin{document}

\newcommand{\rb}[1]{\raisebox{1.5ex}[-1.5ex]{#1}}

\maketitle

\label{firstpage}

\begin{abstract}
We present multi-epoch high-resolution spectroscopy and photoelectric 
polarimetry of the long-period polar (AM Herculis star) \qq. The 
blue emission lines show several distinct components, the sharpest of which
can unequivocally be assigned to the illuminated hemisphere of the secondary 
star and used to trace its orbital motion. This narrow emission line can be 
used in combination 
with \na1-absorption lines from the photosphere of the companion
to build a stable long-term ephemeris for the star:
inferior conjunction of the companion occurs at HJD 
$= 244\,8446.4710(5) + E \times 0\fd15452011(11)$.
The polarization curves are dissimilar at different epochs,
thus supporting the idea of fundamental changes of the accretion geometry, 
e.g.~between one- and two-pole accretion modes.
The linear polarization pulses 
display a random scatter by 0.2 phase units and are
not suitable for the determination of the binary period.
The polarization data suggest that the magnetic (dipolar) axis has a 
co-latitude of $23\degr$, an azimuth of $-50\degr$, and an orbital inclination 
between $50\degr$ and $70\degr$.

Doppler images of blue emission and red absorption lines show a clear 
separation between the illuminated and the non-illuminated hemisphere of the
secondary star. 
The absorption lines on their own can be used to determine the mass ratio 
of the binary by Doppler tomography with an accuracy of $15\% - 20\%$. 
The narrow emission lines of different atomic species show remarkably 
different radial velocity amplitudes: $K_{\rm} = 85 - 130$\,\kmps. 
Emission lines from the most highly ionized species, \hel2, originate 
closest to the inner Lagrangian point $L_1$.
We can discern two kinematic components within  the accretion stream; one is
associated with the ballistic  part, 
the second with the magnetically threaded 
part of the stream. 
The location of the emission component associated with the ballistic 
accretion stream appears displaced between different epochs. 
Whether this displacement
indicates a dislocation of the ballistic stream, e.g.~by a magnetic drag, or
emission from the magnetically threaded part of the stream with 
near-ballistic velocities remains unsolved.
\end{abstract}


\section{Introduction}
QQ Vul was discovered with the HEAO-1 low energy detectors in the $0.15-0.5$
keV band and catalogued by Nugent (1983). Spectroscopic, photometric, and 
polarimetric observations by Nousek et al.~(1984) immediately following its 
detection led to the classification of 
this object as an AM Her type cataclysmic variable 
(or `polar') with an orbital period of 222.5 min. 
Nousek et al.~(1984) found that the centroids of the 
prominent blue emission lines showed a huge systemic velocity,
$\gamma \sim 500$\,\kmps, when 
fitted with a circular orbit velocity equation,
$v = K \sin 2\pi(\phi-\phi_0) + \gamma$.

An extensive observational 
study using X-ray (EXOSAT), UV (IUE), and optical spectroscopy and 
photometry was presented in Osborne et al.~(1986) and Mukai et al.~(1986).
They found pronounced dips in the soft X-ray light curve caused by 
photoelectric absorption in the magnetically confined stream (out of the 
orbital plane). In addition, their spectroscopy revealed a systemic 
velocity $\gamma$ compatible with zero, indicating substantial changes 
in the line emitting region with respect to the Nousek 
et al.~(1984) results. They also identified a narrow emission line component
({\it peak 2}) which they associated with the heated surface of the 
secondary star. 

\begin{table*}
\begin{minipage}{12.2cm}
\caption{
Log of spectroscopic observations of \qq }\label{speclog} 
\begin{tabular}{lllcccr}
\hline
\multicolumn{7}{c}{} \\
\rb{Date} &
\rb{Tel.$^a$} &
\rb{Instr.$^b$} &
\rb{Res.} & 
\rb{$\Delta\lambda$} & 
\rb{T$_{int}$} & 
\rb{No. spec.} \\
\rb{[Y/M/D]} & & & \rb{[{\AA}] FWHM} & \rb{[{\AA}]} & \rb{[sec]} & \\
\hline
86/10/14-20 & LS22 & B\&C+RCA & 3 & 4180--5050 & 600 & 31 \\
86/10/18 & LS22 & B\&C+RCA & 1.5 & 4540--4960 & 625 & 10 \\
88/06/10-12 & CA22 & B\&C+RCA & 3 & 4150--5050 & 720/500 & 22 \\
91/07/08-10 & CA35  & CTS+RCA2  & 1.7 & $4400-4940$ & $660-720$& 38 \\
91/07/08-10 & CA35  & CTS+GEC   & 2.2 & $7550-8800$ & $660-720$& 36 \\
93/08/23-26 & WHT42 & ISIS+Tek1 & 1.5  & $4230-5010$ & 300 & 64 \\
93/08/23-26 & WHT42 & ISIS+EEV3 & 0.7  & $7940-8380$ & 300 & 64 \\ 
\hline
\end{tabular}
$^a$ LS: La Silla, CA: Calar Alto, WHT: William Herschel Telescope 
La Palma; numbers following the observatory code denote the aperture diameter
in units of 10~cm\\
$^b$ instrument codes: B\&C -- Boller \& Chivens 
spectrograph, CTS -- Cassegrain Twin Spectrograph, ISIS -- 
Intermediate-Dispersion Spectroscopic and Imaging System, the
abbreviations following the instrument code denote the CCD chip used
\end{minipage}
\end{table*}

\begin{table*}
\begin{minipage}{10.cm}
\caption[]{
Photoelectric polarimetry of QQ Vul }\label{pollog} 
\begin{tabular}{lccccrc}
\hline
\multicolumn{7}{c}{} \\
\rb{Date} &
\rb{Tel.$^a$} & 
\rb{Instr.$^b$} & 
\rb{Mode$^c$} &
\rb{Filter} &
\rb{T$_{int}$} & \rb{T$_{tot}$} \\
\rb{[Y/M/D]} & & & & & \rb{[sec]} & \rb{[hours]}\\
\hline
85/06/11 & CA22 & ZPP & CP & WL/RG630 & 60 & 3.9 \\
85/06/12 & CA22 & ZPP & LP & WL & 72 & 4.1 \\
87/06/25-27 & CA22 & ZPP & CP & WL & 100 & 8.2 \\
87/09/04 & LS22 & PISCO & LP & WL & 130 & 2.0 \\
88/06/14+16 & CA22 & ZPP & LP & WL & 60 & 5.0 \\
88/06/15 & CA22 & ZPP & CP & WL & 60 & 4.2 \\
\hline
\end{tabular}
$^a$ for coding of telescopes see Tab.~\ref{speclog}\\
$^b$ ZPP: two-channel photometer/polarimeter, PISCO: Polarimeter for 
Instrumental and Sky polarization COmpensation\\
$^c$ CP: circular polarimetry, LP: linear polarimetry
\end{minipage}
\end{table*}

In the same year, McCarthy et al.~(1986) presented their analysis of 
moderate-resolution phase-resolved spectroscopy obtained in a state 
of reduced accretion. They identified a number of narrow emission lines and 
tentatively assigned one of them to the illuminated part of the secondary.
Surprisingly, they found a stationary emission line component at redshift
$\sim$500\,\kmps\ which remained largely unexplained. 

In 1986, Mukai \& Charles detected the secondary star in near-infrared 
spectra of \qq\ and, in 1987, the same authors 
were able to trace the \naI-doublet a\-round the orbital 
cycle. Their data gave the first secure measurement of inferior conjunction 
of the companion star.

EXOSAT observations of \qq\ showed substantial changes of the X-ray light 
curve 
which were tentatively interpreted by Osborne et al.~(1987) as changes between
a simple one-pole and a more complex two-pole geometry. This scenario has 
been supported by Beardmore et al.~(1995) using more recent ROSAT X-ray 
observations. 

The lack of an ephemeris with sufficient accuracy hampers the unique 
interpretation of all the data collected in the past. In this paper, we 
will present an accurate spectroscopic ephemeris using both narrow emission
lines from the illuminated hemisphere of the secondary star and 
photospheric absorption lines observed on several occasions between 1986 and 
1993. We show that both features are equally well
suited for tracing the secondary. Our new spectroscopic ephemeris
replaces older ones which were built using the recorded times of 
linear polarisation pulses or crosscorrelations of optical light curves. 
We will show that the linear polarisation pulses may occur at random 
phase (within a certain interval) and are not suitable, therefore, for the
derivation of the binary period. 

We also present the results of polarimetric observations obtained between 
1985 and 1988. These data are discussed in conjunction with the spectroscopic
observations in order to establish a
consistent picture of the accretion geometry of \qq. Previous polarimetry
was presented by Nousek et al.~(1984) and by Cropper (1998), who derived 
the orbital inclination to be $46\degr < i < 74\degr$ and $i \simeq 
40\degr - 50\degr$, respectively.

This paper concentrates mainly on the 
analysis of the blue emission lines; in the companion paper ({\it 
Mapping the secondary star in \qq}) by Catal\'an, Schwope and Smith (1999, 
henceforth referred to as Paper 1) we present a detailed 
analysis of the absorption and emission lines in the red spectral regime.
Preliminary results were presented in Catal\'an et al.~(1996) and in 
Schwope et al.~(1998, 1999). 

\section{Observations}
\subsection{High-resolution spectroscopy}
\qq\ was observed between 1986 and 1993 using the 2.2-m telescopes at La
Silla and Calar Alto with Boller \& Chivens Cassegrain spectrographs, with
the Calar Alto 3.5-m telescope equipped with the Cassegrain 
double-beam spectrograph TWIN, 
and with the 4.2-m William Herschel telescope (WHT)
at La Palma equipped with the double-beam spectrograph ISIS. All the blue 
spectra covered the approximate
wavelength range 4200--5000\,\AA, where H-Balmer and 
Helium emission lines are prominent. The red spectra were centred on
the \na1-doublet at $\lambda\lambda 8183/8194$\,\AA. Full phase-coverage was 
achieved on all occasions. Some details of the 
spectroscopic observations are given in Table~\ref{speclog}. All observations
were performed under good weather conditions. Flatfield, bias, dark, and
arc line spectra were taken frequently, and spectra of spectrophotometric
standard stars were taken on each of the observing nights. We estimate
the photometric accuracy of our spectra to be better than 30\%.

\subsection{Photoelectric polarimetry}
Polarimetric observations of \qq\ were carried out during several nights in 
1985, 1987 and 1988 using the two-channel photometer/polarimeter of the DSAZ
(Proe\-tel 1978)
mounted on the 2.2m telescope at Calar Alto. In addition, a rather short 
($2^h$) observation covering just one linear polarization pulse was carried
out in November 1987 at the 2.2m telescope at La Silla using the 
ESO-polarimeter PISCO (Stahl et al.~1986). Most observations were done in
unfiltered (white) light, hence the spectral bandpass was limited by the 
atmospheric cutoff and the sensitivity of the photomultipliers 
to $\sim$$3400-8500$\,\AA. An observation log is given in Table~\ref{pollog}.
The sky was monitored regularly, allowing sky-subtraction from object 
measurements after fitting polynomials to the sky measurements.

Using the DSAZ-polarimeter, the incoming signal is modulated by a 
rotating quarter-wave or half-wave plate, for the detection of circular
(CP-mode) and linear polarisation (LP-mode), respectively. In the CP-mode
a linearly polarised signal can also be detected but with reduced efficiency.

\section{Analysis}

\begin{figure}
\psfig{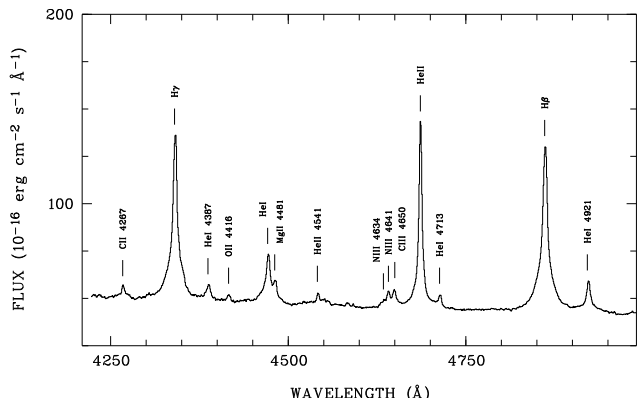}
\caption{Orbital mean blue spectrum of QQ Vul in August 1993}
\label{meansp} 
\end{figure}

\subsection{Spectroscopic ephemeris of the secondary star}\label{s:ephem}
The orbital mean spectrum of our 1993 observations (blue spectral range) is
shown in Fig.~\ref{meansp}. The mean spectra obtained on the other occasions 
look very similar
to that shown in the figure, although at a somewhat different brightness
level. \qq\ displays the typical features of an AM
Herculis star in a high accretion state: a strong line of ionized Helium
\heII, an inverted 
Balmer decrement, asymmetric variable lines with line wings extending 
to $\sim$2000\,\kmps, and a pronounced Bowen blend of C{\sc iii}/N{\sc 
iii}-lines between 4635 and 4650\,\AA. 

\begin{figure*}
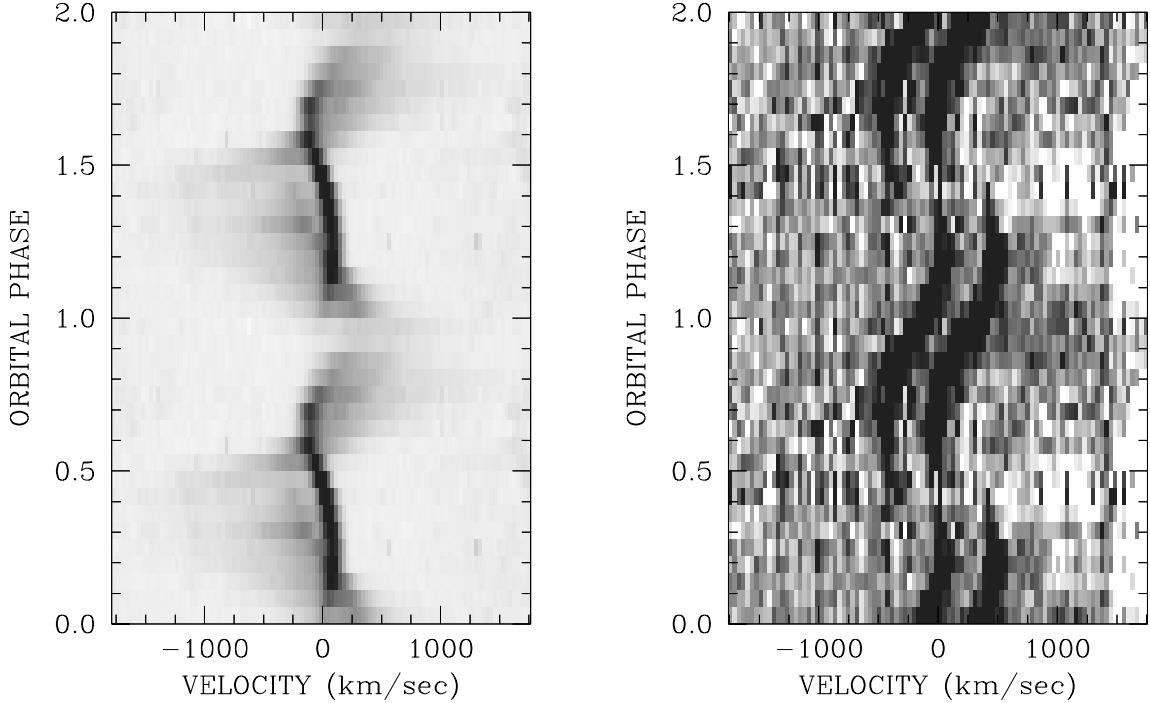

\begin{center}
\parbox[]{70mm}{\
\psfig{figure=lin_he2_91_2p_ps,width=70mm,bbllx=49pt,bblly=190pt,bburx=336pt,bbury=572pt,clip=}
}
\hspace{1cm}
\parbox[]{70mm}{\
\psfig{figure=lin_na1_91_2p_ps,width=70mm,bbllx=49pt,bblly=190pt,bburx=336pt,bbury=572pt,clip=}
}
\end{center}
\caption{Trailed spectrograms of the simultaneously recorded
\heII\ emission and \naI\ absorption lines of \qq\ (1991 observations).
The original spectra were continuum-subtracted and phase-folded 
according to the final ephemeris of Eq.~1. All data are shown 
twice for clarity}
\label{91trails} 
\end{figure*}

Individual emission lines consist of several components with different radial
velocity amplitudes and different brightness variations. To give an example we
show in Fig.\ref{91trails} a grey-scale representation of the phase-folded,
continuum-subtracted  
trailed spectrogram of the \heII-line (1991 observations). The ephemeris used
for phase-binning is that given in Eq.~1. The trailed blue emission line 
spectrogram is shown together with a trailed, continuum-subtracted  
spectrogram of the \na1 absorption line doublet recorded simultaneously. 
These red spectra are fully analysed in the accompanying Paper 1 and are
shown here to facilitate the interpretation of the blue spectra.

\begin{figure}
\psfig{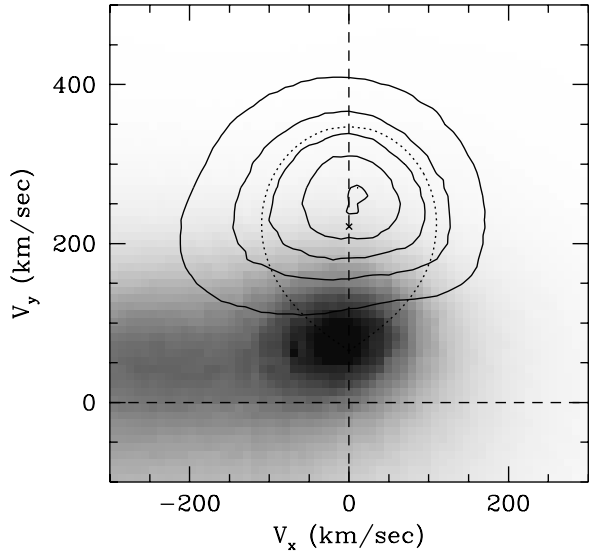}
\caption{Doppler image (MEM-reconstruction) 
of the He\,{\sc ii}\ 
emission and the \na1\ absorption lines recorded simultaneously
in 1991.
The Doppler image of \na1\ is shown as contourplot with 
iso-intensity lines at 20, 40, 60, 80 and 98\% of the maximum intensity.
The Doppler map of  He\,{\sc ii}\ 
is shown as a grey-scale image with dynamic range  
between 0 and 90\% of maximum 
intensity. The overlay gives the size of a Roche lobe for a mass ratio 
$Q = 1.75$, the $\times$ marks the adopted centre of mass of the secondary 
star.
}
\label{na1_he2_91} 
\end{figure}

The profile of the He-line generally shows great similarity to those 
displayed in HU Aqr in its high accretion state 
(Schwope et al.~1997). The most pronounced feature is a narrow emission line 
(NEL) with rather low velocity amplitude. It moves parallel to the 
Na absorption lines, while its brightness is anticorrelated with those of 
the absorption lines. The latter are bright at inferior conjunction
(blue to red zero crossing of the lines), 
the NEL is bright at superior conjunction. The NEL clearly has a much lower 
radial velocity amplitude than the absorption lines. Both the different 
brightness variation and the different radial velocity amplitude
suggest that the two spectral features originate mainly from opposite  
hemispheres of the secondary star. This is demonstrated clearly by a 
combined Doppler map of both lines shown in Fig.~\ref{na1_he2_91}.
The underlying broader features in the He\,{\sc ii}\ emission 
line complex with 
higher radial velocity amplitudes 
must then have their origin somewhere in the accretion 
stream. They will be investigated in section \ref{s:stream}. 
Here we concentrate on the narrow emission line which obviously originates 
somewhere on the donor star.

We cannot confirm the finding by McCarthy et al.~(1986) of a stationary 
narrow emission component at a velocity of $\sim$500\,\kmps.

The \heII\ emission lines in all our spectra obtained during different years 
were fitted by two or three Gaussians in order to separate the NEL from the
stream emission. This was unequivocally possible only in the phase interval
$0.25 - 0.60$, at other phases the NEL is too faint to be clearly detected
or strongly blended with another relatively sharp underlying component 
(around phases $\sim$0.2 and $\sim$0.7). The centroid positions of the NEL
were then fitted assuming a circular orbit velocity equation $v = \gamma + K 
\sin 2\pi(t - t_0)/P$, with fixed period $P$.

\begin{table}
\caption[]{Fit parameters for sine fits to 
the He\,{\sc ii}\  narrow emission and the
\na1 absorption lines}\label{fitparms}
\begin{tabular}{lrr@{$\pm$}lr@{$\pm$}lr@{.}lr}\hline
\multicolumn{9}{c}{} \\
\rb{Year} &
\rb{Cycle} &
\multicolumn{2}{c}{\rb{$\gamma$}} &
\multicolumn{2}{c}{\rb{$K$}} &
\multicolumn{2}{c}{\rb{$t_0$}} & 
\rb{$O - C$} \\
&& \multicolumn{2}{c}{\rb{\kmps}} & \multicolumn{2}{c}{\rb{\kmps}} & 
\multicolumn{2}{c}{\rb{HJD$^a$}} & \rb{days$^b$} \\
\hline
\multicolumn{5}{l}{\it HeII narrow emission line}\\
1986 &$-11183$ & $ -8$ & $11$ & $118$ & $11$ & 6718&4720(20) & $-7$\\
1988 &$ -7266$ & $-66$ & $12$ & $144$ & $13$ & 7323&7330(30) & $+51$\\
1991 &$     0$ & $-13$ & $ 5$ & $116$ & $ 6$ & 8446&4767(31) & $+57$\\
1993 & $  5035$ & $  2$ & $ 8$ & $111$ & $ 5$ & 9224&4817(19) & $+19$\\
\multicolumn{5}{l}{\it NaI absorption line}\\
1985&$-14396$ & $ 17$ & $23$ & $209$ & $28$ & 6222&0000(30) & $+5$\\
1991&$     0$ & $ -3$ & $ 6$ & $271$ & $ 8$ & 8446&47042(62)& $-6$\\
1993&$  5028$ & $ 18$ & $ 7$ & $228$ & $10$ & 9223&3982(11) & $+1$\\
\hline
\multicolumn{5}{l}{$^a +$244\,0000}\\
\multicolumn{5}{l}{$^b \times 10^{-4}$}\\
\end{tabular}
\end{table}

Similarly, the \na1 absorption
lines were fitted by a double Gaussian with a fixed peak separation and the 
same line widths. Again the resulting radial velocities were fitted by
a circular orbit velocity equation and the values given  in Paper 1 are 
quoted here. Ellipsoidal fits gave smaller
residuals than the straight sine fits (Paper 1); the influence on the 
times $t_0$ of zero crossing, however, is negligible. 
We compile the results for  all fit-parameters in Table~\ref{fitparms} and
include also those found in the literature (Mukai \& Charles 1987, \na1
measurement). 

\begin{figure}
\psfig{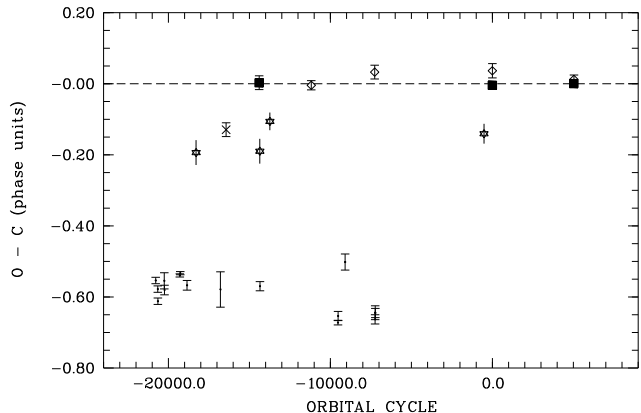}
\caption{Residuals of events around the binary orbit in \qq\ with respect to 
the linear ephemeris of Eq.~1. Meaning of symbols: $\diamond$ 
-- NEL blue-to-red zero crossing; filled square -- Na{\sc i} blue-to-red zero 
crossing; $\times$ -- Mukai et al.~peak `2' component; $\cdot$ -- linear 
polarisation pulses; $\ast$ -- X-ray absorption dips. 
}
\label{omc} 
\end{figure}

Linear regressions for the times of blue-to-red zero 
crossing were performed for the emission and the absorption features. They
yielded identical results within the errors. The weighted 
linear regression for the combined \na1- and He\,{\sc ii}-lines yields

\begin{eqnarray}
T_0 \mbox{(HJD)} & = & 244\,8446.47105 + E \times 0.15452011\\
&& \phantom{244\,8446.47}(48)\hspace{59pt}(11) \nonumber
\end{eqnarray}
(the numbers in brackets are the uncertainties in the last digits). The 
epochs defined by integer numbers of $E$ are regarded as epochs of inferior
conjunction of the secondary star in \qq . All phases in both papers
refer to the ephemeris of Eq.~1. The residuals of several 
features recorded around the orbital cycle, such as 
radial velocity zero crossings, centres of linear polarization 
pulses, X-ray absorption dips, 
are plotted in Fig.~\ref{omc} with respect to Eq.~1. This
diagram  shows the good 
agreement of the narrow emission-line zero crossings with \na1 absorption-line
zero crossings. It shows furthermore that the `peak 2' component of Mukai et 
al.~(1987) which was suspected to originate from the secondary star
has nothing in common with the narrow emission line from the companion 
investigated here. In addition it shows the 
variability of the phase of the X-ray absorption dip (the bars in the diagram
denote the duration of that feature, not the phase error) and 
it shows finally, that the linear
polarisation pulse appears at random phase 
between phase 0.33 and 0.50 and cannot be used 
in order to derive the orbital period.

\subsection{Analysis of polarimetric observations
\label{s:polari}}

\subsubsection{Overall characteristics of the light and polarization curves}

\begin{figure*}
\psfig{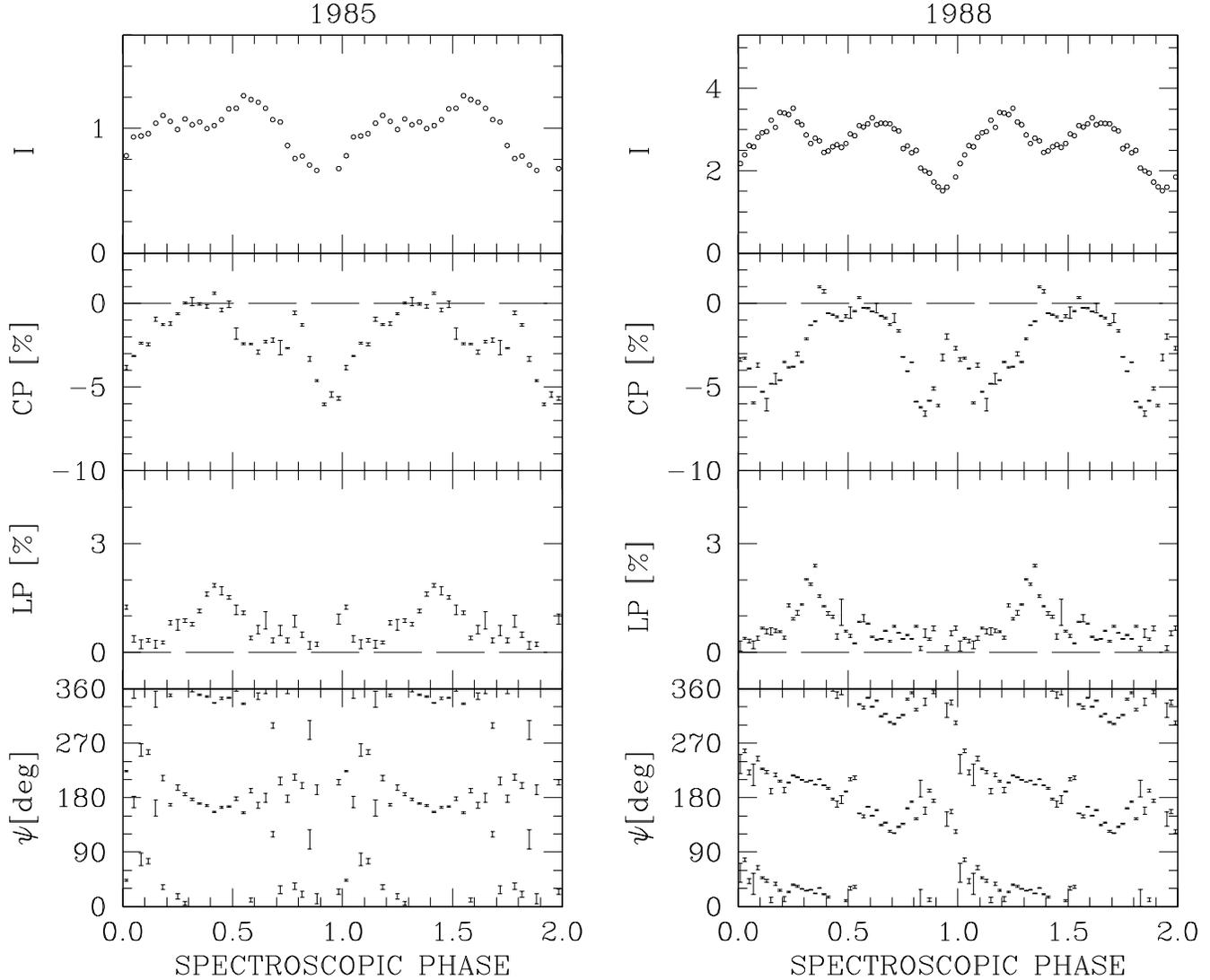}
\caption{
Phase-averaged light and polarization curves of \qq\ obtained in 1985 and 1988
in white light. Shown are from top to bottom the brightness (in arbitrary 
units), the degree of circular and linear polarization and the polarization 
angle. The data are shown twice for clarity. Phase zero is
the inferior conjunction of the secondary star (ephemeris given in 
Eq.~1).
}
\label{pol8588} 
\end{figure*}
In Fig.~\ref{pol8588} we show phase-averaged light and polarization curves
with full phase coverage obtained in 1985 and 1988 in white light. All data 
were phase-folded and averaged using the ephemeris derived in 
Sect.~\ref{s:ephem}, where phase zero indicates inferior conjunction of the 
secondary star.
The data obtained in 1987 are very similar to those obtained one year 
later, but with poorer signal-to-noise. These data are therefore not discussed
separately.

The 1985 data are quite similar to those published recently by Cropper (1998)
and to those by Nousek et al.~(1984)
but show some striking differences from the 1988 data with respect to
the shape of the curves and the phasing of individual features.

The light curve at all three epochs is 
double-humped with primary and secondary 
minima separated by a half orbital cycle and the primary minimum centred
around $\fisp \simeq 0.9$. In June 1985, 
maximum light occurred at the 
phase of the second hump whereas in all later data 
the first hump was stronger (see Fig.~\ref{f:3lcs}).
The phasing of the light curve does not show dramatic
changes between 1985 and 1988, 
quite unlike the circular and linear polarisation curves which do.

Circular polarization  is almost always negative with short excursions at 
low level towards positive values somewhere in the phase interval $\fisp = 0.30
- 0.55$, and shows a pronounced depression at phase $\fisp = 0.80$ and 
$\fisp = 0.95$ in 1985 and 1988, respectively.

Linear polarization is always low, i.e.~below 2--3\%. The curves show one
distinct linear polarization maximum (pulse) per orbital cycle. It is centred 
at phase $\fisp \simeq 0.42$ in 1985, and marks the end of the 
phase interval of vanishing circular polarization in that year. In 1988, the 
pulse occurs at $\fisp \simeq 0.35$, and marks the start 
of the interval of positive (or vanishing) circular polarization.
Not so easily recognizable in the phase-averaged representation of 
Fig.~\ref{pol8588} is a second fainter linear polarization pulse at 
$\fisp = 0.57$ observed in 1988 
which coincides with a second circular polarization zero 
crossing.
The linear polarization angle displays a large scatter in 
1985, whereas 
the variation is much simpler in 1988. Meggitt \& Wickramasinghe (1982)
derived a simple relation for the derivative of the polarization angle
at the time of the polarization pulse 
for the case of a point-like emission region, $\dot\psi_{\rm p} = 
\cos{i}$. The observed data do not allow us to determine a
unique value for $\dot\psi_{\rm p}$ in 1985, contrary to 1988 where $i
\simeq 60\degr$. The fact that the value of $\dot\psi_{\rm p}$ 
changes during the pulse in the 1985 observation
means that the emission region must have had a more complicated shape 
on that occasion compared with the 1988 observation.

The optical light and polarization curves of AM Herculis binaries are
modulated by cyclotron beaming, eclipses by the secondary
star and the accretion stream, and self-eclipses of the accretion region by 
the white dwarf itself. X-ray observations on different occasions using 
the EXOSAT satellite have shown that there is no eclipse by the secondary 
star (Osborne et al.~1987). The polarimetric data in 1985 were taken 
simultaneously with an EXOSAT observation. The combined data show
clearly that the depression in the circular polarization at phase
$\fisp \simeq 0.8$ is due to an obscuration of the accretion spot by the 
intervening accretion stream. It is that part of the stream  
which is magnetically coupled and lifted out of the 
orbital plane. As has been pointed out by Cropper (1998), this 
requires the orbital inclination $i$ to be larger than the colatitude of 
the accretion spot $\delta_{\rm s}$ (measured with respect to the rotation 
axis). 

Around the phase of the X-ray and circular polarization dip the observer 
looks almost directly onto the accretion region (minimum angle between
the line of sight and the surface normal at the accretion spot). At this
viewing angle the optical flux and the circular polarization are reduced due
to strong cyclotron beaming. The primary optical minimum (and circular 
polarization dip) is, therefore, caused by the 
combination of depolarization/absorption 
in the stream and cyclotron beaming. 

The secondary optical minimum at phase $\fisp \simeq 0.4$ marks the phase 
when the accretion spot is at the limb of the white dwarf. 
Its phasing coincides with a sign reversal of circular polarisation 
and the occurrence of linear polarisation pulses. On the basis of those
clues, we interpret the 
pronounced secondary minimum in the light curve 
seen in 1988, which is even more pronounced in the 
1991 and 1993 data (see Fig.~\ref{f:3lcs}),
as caused by a partial self-eclipse of an extended accretion region. The 
EXOSAT X-ray light curve of September 14/15 1985 shows a broad primary minimum 
at that phase which is being interpreted by us as the combined effect of 
the partial self-eclipse and foreshortening 
of an extended  accretion spot.

\begin{figure}
\psfig{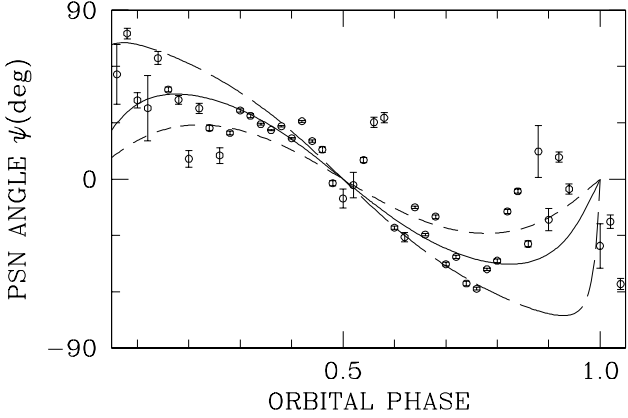}
\caption{Variation of the linear polarization position angle 
in the 1988 data shown together with some models. The three model 
curves correspond to $i = 70\degr$ (short-dashed line), 60\degr
(solid line), and 50\degr (long-dashed line) and length of self-eclipse
of 0.21 phase units.
}
\label{psn_ang} 
\end{figure}

\subsubsection{Modelling the 1988 polarimetry}

\begin{figure}
\begin{center}
\parbox[]{64mm}{\
\psfig{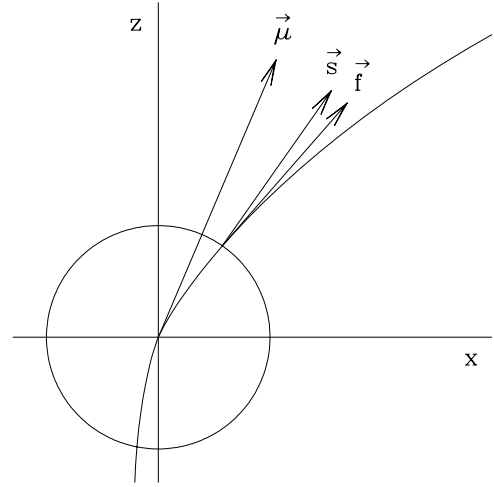}
}
\end{center}
\caption{Sketch of the dipolar geometry at the white dwarf surface. Shown are
a cut through the white dwarf and a single dipolar field line projected on
an appropriately rotated $x'-z$ plane. Also shown are
vectors indicating the orientation of the dipolar axis $\vec{\mu}$, of the
surface normal in the accretion spot $\vec{s}$, and of the magnetic field 
line in the accretion spot $\vec{f}$, the latter being tangent to the 
field line at the footpoint. 
In a dipolar geometry $\vec{s}$ and $\vec{f}$ are always different.
}
\label{f:dip_geo} 
\end{figure}

\begin{figure}
\psfig{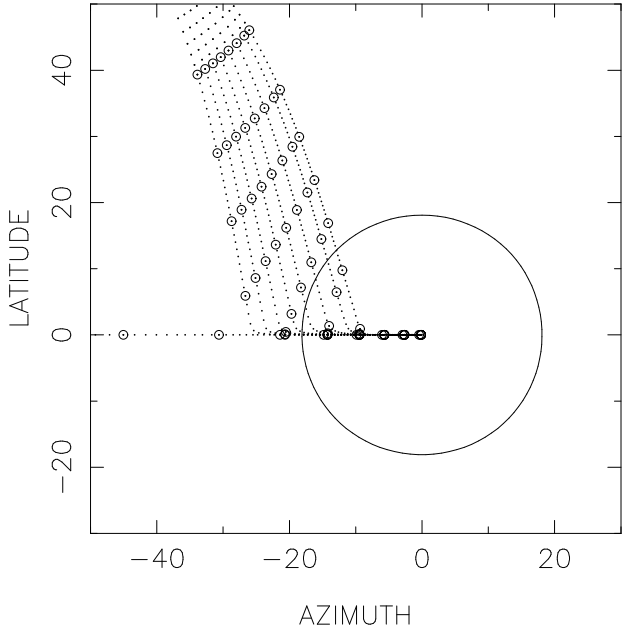}
\caption{Model for the coupling of the accretion stream as seen 
by a hypothetical observer located at the  white dwarf. The big
circle represents the secondary star. The adopted 
orientation of the dipole $\vec{\mu}$ is given by its azimuth $\chi_\mu = 
-50\degr$ and its colatitude $\delta_\mu = 23\degr$.
}
\label{f:viewwd} 
\end{figure}

To determine the accretion geometry of \qq\ we take the simple 
view of a pointlike accretion region in a dipolar geometry. A simplified 
sketch of the geometry is shown in Fig.~\ref{f:dip_geo}. Three vectors play 
an important role, the magnetic axis $\vec{\mu}$, the surface normal in the 
accretion spot $\vec{s}$, and the field in the spot $\vec{f}$. All vectors
are shown projected onto an $x'-z$ plane centred on the white dwarf with 
$z$ being parallel to the rotation axis. 
For an analogous 
representation of the geometry see e.g.~Cropper (1989). The stellar
co-latitudes of the dipolar axis, the spot and the field therein are 
measured with respect to the rotation axis and designated $\delta$ with the
corresponding subscripts $\mu, s, f$. Similarly, the azimuths, $\chi$, 
of these vectors are measured with respect to the line joining both stars.
Conventionally the accretion stream in our geometry 
is deflected towards negative azimuth. Below we will discuss  also the azimuth 
of the threading region, $\chi_{\rm th}$ (Fig.~\ref{f:viewwd}), which 
is defined in the same reference frame.
 
Previous determinations of the accretion geometry based on polarimetry
were published by Nousek et al.~(1984) and 
Cropper (1998). Nousek et al.~used a
point-like accretion spot and derived a stellar latitude of the accreting
magnetic pole in the range $27\degr - 10\degr$, and an orbital inclination 
of $i = 46\degr - 74\degr$. Cropper (1998) investigated extended accretion 
arcs with constant temperature, magnetic field and density, and derived
$i \simeq 40\degr - 50\degr$ and a co-latitude of the magnetic dipolar axis
$\delta_{\mu} \simeq 10\degr$. 

The markedly different yet simpler behaviour seen in the 1988 data
relative to all previous observations, called for a redetermination
of the accretion geometry. Furthermore, with our extended data set, 
it is possible to find a solution that can encompass the results 
from both polarimetry and Doppler tomography. 
Previous analyses also lacked an absolute reference
in the binary frame. We do not attempt to perform 
a least-squares fit of the polarimetric data, instead we try
to globally optimize the  model (including polarimetry, X-ray photometry 
and Doppler tomography).
Cropper (1998), although using a more sophisticated model which accounts
for an extended accretion arc, found a full inversion of the data 
(as in Potter, Hakala \& Cropper 1987) objectively inefficacious and 
also only globally optimized his model.

The following observed features are used for an estimate of the geometry: 
the length of the self-eclipse $\Delta\phi_{\rm self}$, the variation 
of the polarization 
angle, and the phase of the circular polarization dip. Variation of the 
dipolar axis 
$\vec{\mu}$, the inclination $i$, and the coupling radius then 
predicts a different run of the polarization angle. By integrating 
the equation of motion for one particle with given mass ratio $Q$, one can
also predict its velocity pattern in the $(v_x,v_y)$-plane (a schematic 
Doppler map) when projected with the assumed orbital inclination.
The resulting image can then be compared with the observed 
Doppler maps (Sect.~\ref{s:stream}) thus constraining the azimuth 
$\delta_{\rm th}$ of the threading region.
A solution is accepted if the predicted tomogram and the predicted curve of
polarization angle are globally optimized and the phase of the 
X-ray/circular polarization dip is reflected well.
 
The separation between the linear polarization pulses is 0.24 phase units.
The sign reversals of the circular polarization are 
separated by 0.18 phase units. For our estimate we use the mean of both 
values, $\Delta\phi_{\rm self} = 0.21$. 
Both features are, in principle,
measures of the same event, namely when the line of sight is perpendicular 
to $\vec{f}$. The length of the self-eclipse of the accretion spot
is shorter than this phase interval.
The inclination is approximately 60\degr, as suggested by the slope
$\dot\psi_{\rm p}$ of the polarization angle during the pulse 
(Fig.~\ref{psn_ang}). The 
values of $i$ and $\Delta\phi_{\rm self}$ constrain the possible values of
the colatitude $\delta_{\rm f}$ of the field line to a certain
range (see Brainerd \& Lamb, 1985, for technical details).
The dip is centred at $\fisp = 0.95$, corresponding to an azimuth of 
about $-20\degr$.

Model curves for the polarization angle for the acceptable range of inclination
angles, $i = 50\degr - 70\degr$, are shown in Fig.~\ref{psn_ang} together
with the 1988  data. Deviations between model and
observation in the phase interval 0.8 -- 1.1 should not be taken seriously,
since this is the phase of depolarization in the stream and low intensity
due to cyclotron beaming. The adopted values of the magnetic colatitude 
for the three model curves are $\delta_{\rm f} = 47\degr$, $38\degr$
and $27\degr$ for $i=50\degr$, $60\degr$ and $70\degr$, respectively. 

A globally optimised fit taking into account the different observational
quantities was achieved with the following 
system parameters: mass ratio $Q = M_1/M_2 = 1.75$, inclination $i = 60\degr$,
orientation of the dipole axis 
$\delta_\mu = 23\degr$ and $\chi_\mu = -50\degr$, 
azimuth of the threading region $\chi_{\rm th} = -20\degr$. 
With these parameters the accretion spot emerges at 
$\delta_s = 33\degr$ and $\chi_s = -40\degr$, and the field line within 
the spot has 
$\delta_f = 36\degr$ and $\chi_f = -37\degr$. In Fig.~\ref{f:viewwd} a series
of field lines around this particular one is shown as seen from the white 
dwarf. An observer viewing the binary at inclination $i$ would 
be rotating along a line of constant latitude $l = 90\degr - i$.
Hence, the figure illustrates that the line of sight for a likely 
inclination of about $60\degr$ crosses the stream between $\chi = -17\degr 
\mathrm{\dots} -32\degr$, which corresponds to orbital phase 
$\fisp = 0.91\mathrm{\dots} 0.95 $. At that phase the X-ray absorption dip 
was observed in the September 1985 data.
The model thus reproduces the observed motion of the polarization 
angle, the observed phase of the X-ray/circular polarization dip, 
and it reflects the main features of the Doppler tomograms 
(see Sect.~\ref{s:stream}). 

Based on the analysis of the 1988 polarimetry we do not claim an 
accuracy of the orbital inclination $i$ and the co-latitude $\delta_f$
better than $10\degr$. One should bear in mind, however, that these 
two quantities cannot be varied independently, they communicate with 
each other via $\Delta\phi_{\rm self}$. A higher inclination requires
a lower co-latitude of the field $\delta_f$ and of the spot $\delta_s$.
We note in particular that our estimate of $i$ based on polarimetry is
in full agreement with the results of Paper 1, $i = 65\degr$, using 
high-resolution spectroscopy of the infrared Na lines. It is also 
in agreement with the analysis in Nousek et al.~(1984), which is worth 
mentioning since their data look quite different. The somewhat 
lower values favoured by Cropper (1988), $i = 40\degr - 50\degr$, are 
likely to be caused by the large scatter of the polarization angle 
in his data which indicates a more complicated accretion geometry
in 1983 -- 1985.

Our results imply certain values 
of the azimuth and co-latitude of the accretion spot, which are 
in good agreement with X-ray photometry (not used for the determination 
of the geometry so far). The value of $\chi_s = -40\degr$
predicts a minimum in the X-ray flux at phase 0.39 due to projection effects.
Indeed, this is exactly the phase where the broad 
X-ray minimum occurred in the EXOSAT data taken in September 1985
(Osborne et al.~1987).
Although the X-ray and our polarimetric observations were not carried 
out simultaneously, the simple
shape of the X-ray light curve in 1985 and the simple polarimetric
behaviour of \qq\ seen in 1988 are suggestive of a similar accretion geometry.
On the other hand, 
our parameter combination of $i$ and $\delta_s$ 
predicts a complete self-eclipse of
the point-like accretion region at that phase, 
$i + \delta_s = 93\degr > 90\degr$,
which is not observed. However, this does not necessarily imply that the 
model is incorrect. For one, the uncertainties associated with our 
derived values are larger than $5\degr$. More importantly, 
the accretion region is likely to be extended.

\subsubsection{Changes of the accretion geometry}

Up to this point, we have considered and modelled only the 1988 observations.
The shift  of the X-ray absorption dip towards earlier phases in June 1985 
and October 1983, 
when the X-ray light curve had a complex shape, suggests that
threading of the stream occurs further downstream, i.e.~farther 
away from the $L_1$--point at an 
azimuth of $\chi_{\rm th} \sim -70\degr$. The X-ray lightcurve 
of June 1985, on the other hand,
shows a completely different shape from the September 1985 data,
which cannot be explained
by a simple migration of the accretion spot towards a different azimuth.
The fact that phasing of the X-ray maximum in June 1985 occurs 
exactly at the phase of minimum flux in the light curve taken three months
later suggests the presence of a second 
active accretion region on the opposite hemisphere (Osborne et al.~1987).
The corresponding cyclotron emission 
from the second region  must affect the polarization measurements
and probably distorts the simple variation of the polarization angle 
seen in 1988, which likely represents a one-pole accretion mode.
This was already pointed out by Cropper (1998) who also detected a second
polarization pulse at optical minimum phase. We confirm this likely 
presence of a second accretion region based on our  observation of 
enhanced linear polarization at $\fisp = 0.0$ in our 1985 data.

\subsection{Analysis of the emission lines}

\subsubsection{Origin of the narrow emission lines}
\label{s:repro}

\begin{figure}
\psfig{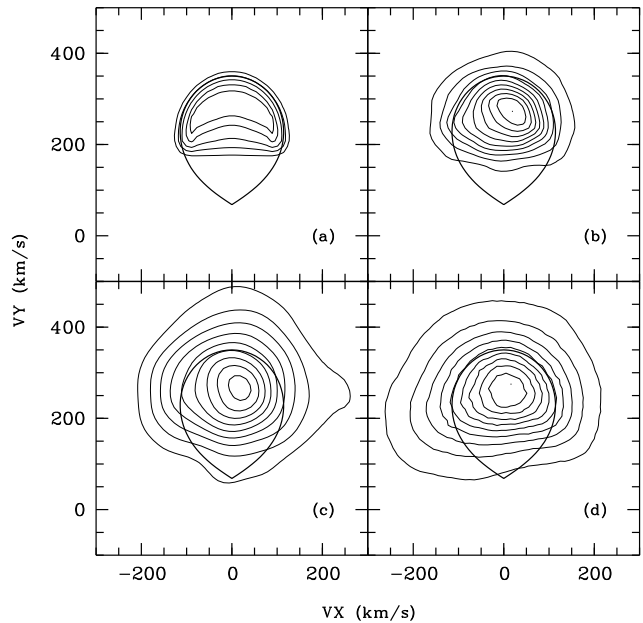}
\caption{Numerical experiments showing the effects of noise and sampling 
on the reconstructed Roche lobe. The input Roche lobe used for generation 
of synthetic spectra is shown in each panel. 
{\it (a)} 
Reconstruction without noise, infinite velocity resolution and velocity
sampling 12\,\kmps, phase sampling 0.00556;
{\it (b)} 
reconstruction with noise at observed level, velocity resolution 80\,\kmps, 
phase sampling 0.0556;
{\it (c)} 
reconstruction with noise at observed level, velocity resolution 180\,\kmps, 
phase sampling 0.0556;
{\it (d)} reconstruction of the observed (1991) spectra. 
In panels (b) -- (d) the isocontours represent between 10\% and 90\% 
of maximum intensity in steps of 
20\%. The isocontours drawn in (a) are the 20\% -- 80\% intensity 
levels in steps of 20\% with an additional contour at 90\%.
}
\label{f:na_sim} 
\end{figure}

Our initial analysis of the simultaneously recorded spectrograms of the 
He\,{\sc ii}\ emission and the \na1\ absorption lines in Sect.~\ref{s:ephem}
based on radial velocity fits suggests that the NEL and the absorption
lines originate on opposite hemispheres of the secondary star. This impression
is supported by a combined Doppler tomogram of both lines 
(Fig.~\ref{na1_he2_91}) which was constructed using 
MEM-deconvolution of the trailed spectrograms. We had chosen the \hel2\
emission line for this initial analysis because it is intrinsically 
the sharpest of the bright lines in the blue spectral regime. With 
a full width at half maximum FWHM of 1.9\,\AA\ (1991) and 2.0\,\AA\ (1993)
at phase 0.5, the NEL of \hel2\ is just resolved. The NEL components of the
H-Balmer lines at the same phase have a FWHM of 3.7\,\AA , those of \he1\
have 2.5 -- 4\,\AA. The NEL of \mgII\ is comparable in width to that in \heII\
but  has a much lower  flux and thus reduced signal-to-noise. 

A comparison of the results from the fit to the radial velocity measurements
and those from the Doppler map (Sect.~\ref{s:ephem}) for the same data
set reveals an inconsistency
between the radial velocity amplitudes in the NEL.
Maximum emission in the Doppler map occurs at $v = |v_y| \simeq 96$\,\kmps\
whereas the radial velocity fit gives $v = 116\pm6$\,\kmps. We think
that this inconsistency results from the reduced phase interval in which the 
NEL velocities can be derived with good accuracy.
Thus, we will use radial velocity amplitudes derived from the 
Doppler maps which are based on all the data instead of a restricted data
set. To compute  the Doppler maps we mainly used the code
of Spruit (1998). 

\begin{figure}
\psfig{figure=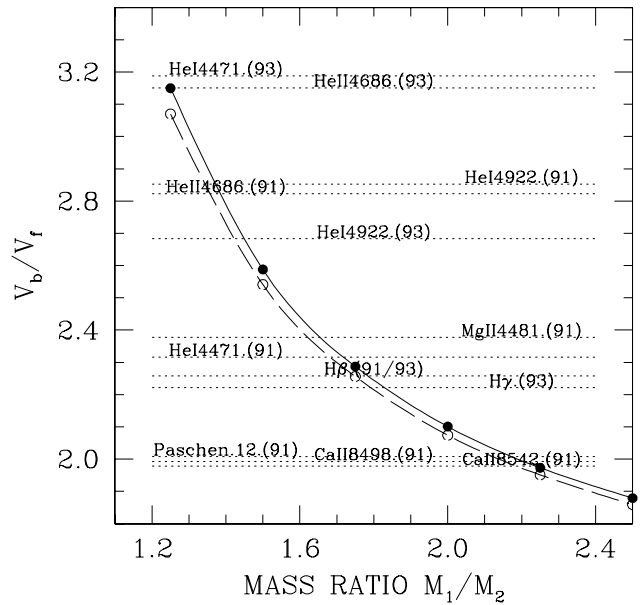,width=84mm,bbllx=87pt,bblly=230pt,%
bburx=516pt,bbury=647pt,clip=}
\caption{Velocity ratio of the photocentres of the irradiated front- and the
non-irradiated backside of the donor star due to our irradiation model 
compared with the observed velocity ratios of \na1\ absorption 
lines (271\,\kmps) 
and narrow emission lines (NEL) of different species (data from 
1991 and 1993). The solid line is a spline fit through the filled dots 
which were computed for $i=90\degr$, the open circles were computed for 
$i=60\degr$ (fit indicated by dashed line).
}
\label{f:q_vrat} 
\end{figure}

The relatively sharp separation between emission and
absorption features in the Doppler map shown in Fig.~\ref{na1_he2_91}
requires us to  address  the following questions. (1) Is it possible
to reconstruct the shape of the Roche lobe and thus determine the mass 
ratio $Q$ using Doppler maps of the \na1-lines alone?
(2) Is it possible to measure the size of the 
Roche lobe of the secondary directly using the velocities of the illuminated
and the non-illuminated hemispheres of the secondary star provided these
could be unequivocally determined observationally? Answering these
questions, in particular locating the $L_1$-point, is essential 
for the next section's investigation of the accretion stream using Doppler 
tomography.


\subsubsection{An irradiation model for the secondary star}

To address both questions we computed synthetic line profiles
using an illumination model of the secondary star like the one used 
in MR~Ser and HU~Aqr (Schwope et al.~1993, Schwope et al.~1997). The 
procedure is described in detail in  Beuermann \& Thomas (1990). 

In brief, the irradiating source is located at $M_1$, it is assumed that the 
secondary star fills its Roche lobe, that no shielding of the secondary's
surface by  the accretion stream or an accretion curtain occurs and that 
a sharp separation of irradiated  and non-irradiated parts of the 
secondary exists. We assume that both emission and absorption lines 
originate from the photosphere, i.e.~from the Roche lobe of the secondary.
The shape of the Roche lobe is iterated for given $Q$. 
Line emission from a surface element on the irradiated secondary is taken to 
be proportional to the solid angle as seen from the source at $M_1$. 
Radiation is assumed to be optically thick, hence foreshortening 
of surface elements has to be taken into account.
For the non-illuminated parts  of the secondary star we assume
that an absorption line is formed whose strength is proportional 
to the size of the surface element and also take foreshortening of the 
individual element into account.
Emission and absorption line profiles are
synthesized using the full Roche geometry. Since the spread of 
the radial velocity over the secondary's surface is of the order 
of several hundred \kmps, we regard velocity broadening as the only 
broadening mechanism of spectral lines. 
In order to transform velocities from dimensionless 
binary coordinates into the observer's 
frame, we adopt the mass-radius relation by Neece (1984) for the secondary 
star. The often used
empirical relation by Caillault \& Patterson (1990) gives slightly different 
results (see below).

As mentioned above, 
our model does not take into account any kind of shielding of
the secondary star by e.g.~an accretion curtain. This was present at some 
occasions in AM~Her and HU~Aqr (Davey \& Smith 1996, Schwope et al.~1997)
and became obvious in these stars 
by a prominent left/right asymmetry of the Doppler maps.
Such an obvious asymmetry is not present in the combined He/Na Doppler map
of \qq, the centres of light of both species lie at 
$v_x = 0$\,\kmps\ (Fig.~\ref{na1_he2_91}), 
i.e.~indicating no or far less shielding in 1991 than in 1993.
Shielding was present in 1993 (cf.~Paper 1), with the effect that the radial
velocity curve of \na1\ had a higher ellipticity, yet smaller
amplitude. 

In order to answer question (1) we computed absorption line 
profiles with the same spectral resolution, orbital phase sampling and noise
level as our 1991 observations. 
In Fig.~\ref{f:na_sim} we compare reconstructions of data with 
(a) infinite resolution and S/N, (b) resolution, sampling and noise as
in 1991 ($v_{\rm FWHM} = 80$\,\kmps\ measured using arc lines), 
(c) sampling and noise as before but degraded resolution 
($v_{\rm FWHM} = 180$\,\kmps) and (d) with a reconstruction of the observed 
data. The Roche lobe used for generation of the data is shown in each 
panel. 

From that exercise we learned (panel a) that 
it is possible to reconstruct the original shape of the
non-irradiated hemisphere of the companion star using 
high-quality data (high resolution, low noise), not a big surprise. 
The reconstruction shown in panel (b),
which uses data of same quality as the observed data
appears somewhat smeared. This map has already lost the arc shape of the 
inner contours of the reconstruction in panel (a). 

Due to the limited resolution of 
the data, the contours extend into the nominally dark, 
front-side hemisphere of the star. The whole map has a  
smaller FWHM than the reconstruction of the observed data (panel d).
In order to match the FWHM of the map in panel (d) we had to degrade 
the resolution in our simulation to values as high as 
$\Delta v = 180$\,\kmps. The resulting map (c) is almost rotationally
symmetric and  no longer displays  a half star as all other maps do.
The high degree of similarity between maps (b) and (d) suggest that the
\na1\ absorption almost entirely originates from the non-irradiated hemisphere
of the companion star and that the size of the Roche lobe chosen for the 
simulations is a fairly good representation of the true size. 
Variation of $Q$ increases the size and shifts the lobe 
along the $v_y$-axis and makes the fit worse.

Our guess is that the 
observed map is disturbed by some effect and is, therefore,  not as sharply 
reconstructed as the simulated one. For example, emission 
from the nearby He\,{\sc ii}\,$\lambda$8236 line and the corresponding curved
continuum could easily contaminate the absorption lines. 
The range of $Q$, which gave acceptable fits (by eye)
to the location and curvature of the contours of the observed
Doppler map, is $Q = 1.5 - 2.2$ (for $i$ between 50\degr\ and 71\degr)
with a best agreement at about $Q = 1.8$.

Figure \ref{f:q_vrat} addresses question 2. We attempt 
to (further) constrain the mass ratio using the velocity ratio of the 
photocentres of the illuminated and non-illuminated 
hemispheres of the secondary star. Synthetic spectra computed 
for different mass ratio $Q$ and orbital inclination $i$ were used.
The inclination only has a small effect on the velocity ratio. 
Theoretically, the velocity ratio $v_{\rm b}/v_{\rm f}$ is 
a steep function of $Q$. This -- in principle -- offers the opportunity to
measure the size of the Roche lobe by determining the velocities 
of the photocentres. We assume that, observationally, the velocity of the 
non-irradiated side of the secondary star is $v_b =271$\,\kmps, the
largest velocity amplitude of the \na1\ lines measured so far (1991 data).
Above we have argued that the Doppler map is
consistent with a complete depletion of the irradiated front side 
of Na absorption. Incomplete Na depletion would mean that the true
$v_b$ would be larger and the derived mass ratio be smaller.

Using all NEL's radial velocity amplitudes with 
estimated uncertainties better than 10\,\kmps\ as 
potential tracers of $v_f$, observed ratios $v_b/v_f$  were computed.
These ratios, shown as dotted lines in Fig.~\ref{f:q_vrat} are highly 
dispersed, from 85\,\kmps\ for He{\sc i}\,4471 in 1993 to
135\,\kmps\ for Ca{\sc ii} in 1991, 
and do not allow us to  constrain the mass ratio better than when
using the \na1\ lines alone (see above and Paper 1). 

For $Q \simeq 1.85$ and $i = 65\degr$ (Paper 1), the predicted velocity 
$v_f$ is $115 - 125$\,\kmps, 
that of the $L_1$-point is $v_{\rm L1} \simeq 75 - 80$\,\kmps,
$v_b/v_f =2.36-2.17$. 
For these parameters the H-Balmer lines are apparently the most 
suitable tracers of $v_f$.
The Ca and Paschen lines are compatible with this solution within 
the accuracy of our velocity determination, while the \hel2\ lines
display too low velocities, i.e.~they originate closer to the $L_1$-point.
Obviously these lines are formed in an extended 
quasi-chromosphere located above the photosphere. The He ionizing 
radiation is not able to reach the whole geometrically allowed hemisphere 
of the secondary. The data
presented here are thus able to resolve (although marginally) these different
layers. A quantitative understanding of the formation of the 
emission lines requires detailed modelling
of non-isotropically X-ray irradiated photospheres of late-type stars, 
a project within reach of current models (e.g.~Hauschildt, 
Baron \& Allard 1997), but not yet undertaken. 

\subsubsection{Multi-epoch line flux variations}

\begin{figure*}
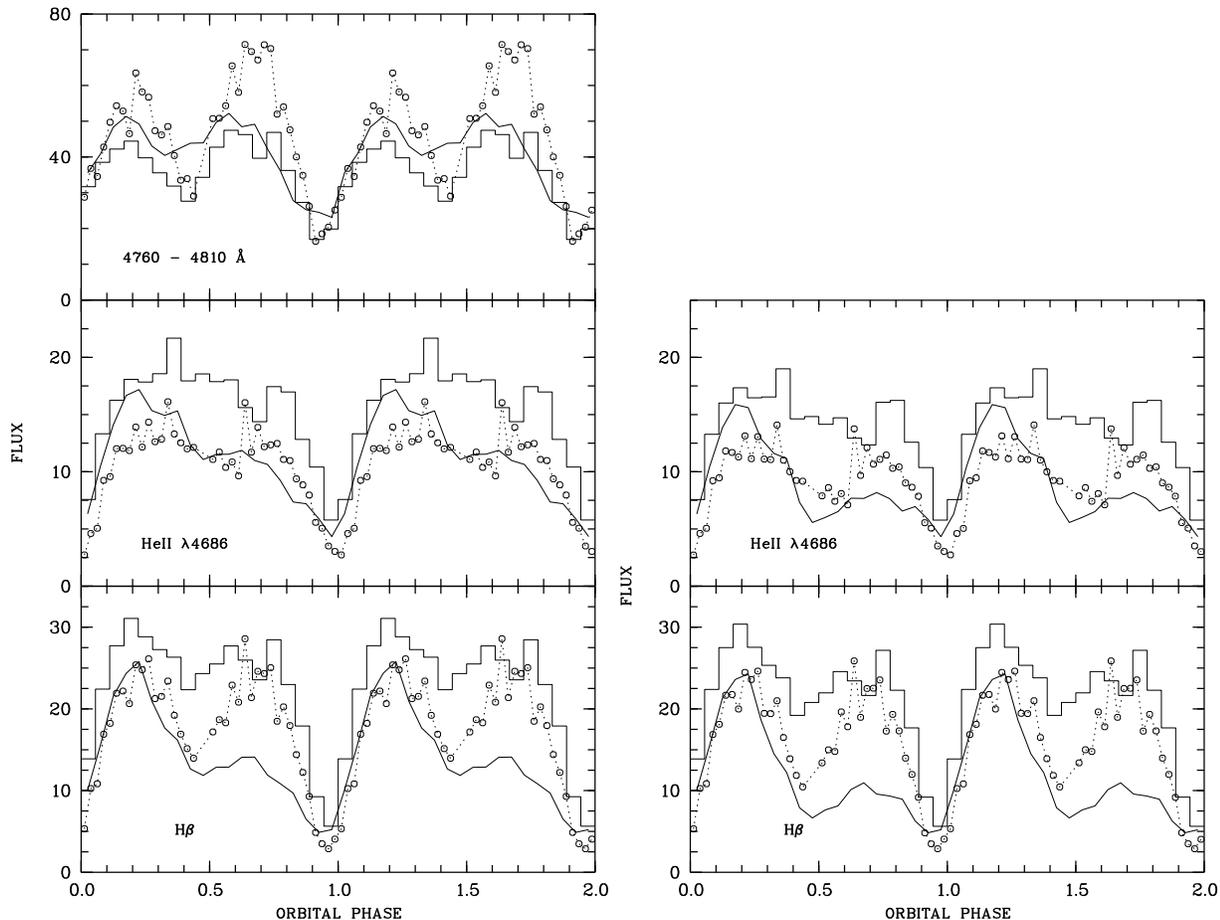

\begin{minipage}{84mm}
\psfig{figure=3lcs_ps,width=80mm,bbllx=60pt,bblly=84pt,%
bburx=510pt,bbury=770pt,clip=}
\end{minipage}
\begin{minipage}{84mm}
\psfig{figure=3lcs_stream_ps,width=80mm,bbllx=60pt,bblly=84pt,%
bburx=510pt,bbury=770pt,clip=}
\end{minipage}
\caption{
{\it (a, left)} Continuum and emission line light curves (\heII, H$\beta$) of
\qq\ in 1986 (solid lines), 1991
(histograms), and 1993 (circles). The continuum flux is the 
wavelength-averaged flux over the line-free wavelength interval indicated 
in the top panel in units of 10$^{-16}$\,erg\,cm$^{-2}$\,s$^{-1}$\,\AA$^{-1}$.
Line flux is the integrated line flux in units of 
10$^{-14}$\,erg\,cm$^{-2}$\,s$^{-1}$. 
{\it (b, right)} Emission line light curves after subtraction of the 
approximated contribution of reprocessed emission from the secondary star
thus leaving emission only from the accretion stream. 
}
\label{f:3lcs} 
\end{figure*} 

For the analysis of emission lines originating from the stream, we make 
use of emission line light curves and Doppler maps, shown in 
Figs.~\ref{f:3lcs} and \ref{f:3mems}. The emission line light curves 
are based on the integrated line profiles of the trailed spectrograms 
shown in Fig.~\ref{f:3mems}. We approximately removed the contribution 
of the NEL, which does not originate from the stream, by measuring its
flux at phase $\fisp = 0.5$, adapting a simulated emission line light 
curve to that level and subtracting it from the total flux. These NEL-free
light curves are also shown in Fig.~\ref{f:3lcs}. For comparison, continuum 
light curves extracted from a line-free 
region between the two emission lines
H$\beta$ and \heII\  are also shown. 

The continuum light curves are all similar in shape displaying a double-humped
structure with the primary minimum centred on $\fisp \simeq 0.95$ in 1991 and
1993 and probably somewhat earlier, $\fisp \simeq 0.92$ in 1986. 
As discussed in Sect.~\ref{s:polari} this minimum is caused by cyclotron 
beaming and the small shift of minimum phase at the  different epochs
indicates a possible azimuthal shift of the emission region. 

The emission line light curves are also double humped with intensity maxima at 
phase 0.2 and 0.7. This behaviour is more pronounced in H$\beta$ 
than in \hel2\  
and more obvious  after subtraction of the NEL, 
which peaks at $\fisp = 0.5$. 
Pronounced minima occur in the phase interval 0.92 -- 1.00, the minima
in H$\beta$ generally occur earlier by $\sim 0.03 - 0.05$ phase units at all 
epochs. 
The minima in both lines occurred earlier in 1986 than in 1993.
The light curves in 1986 are highly asymmetric with a pronounced first 
maximum which is more than double the height of the second one.
All these observed properties are suggestive of optically 
thick radiation in the 
accretion stream, with the stream having a higher thickness in 1986 than 
at the two other epochs.

\subsubsection{Multi-epoch trailed spectrograms and Doppler maps of \heII }
\label{s:stream}

\begin{figure*}
\psfig{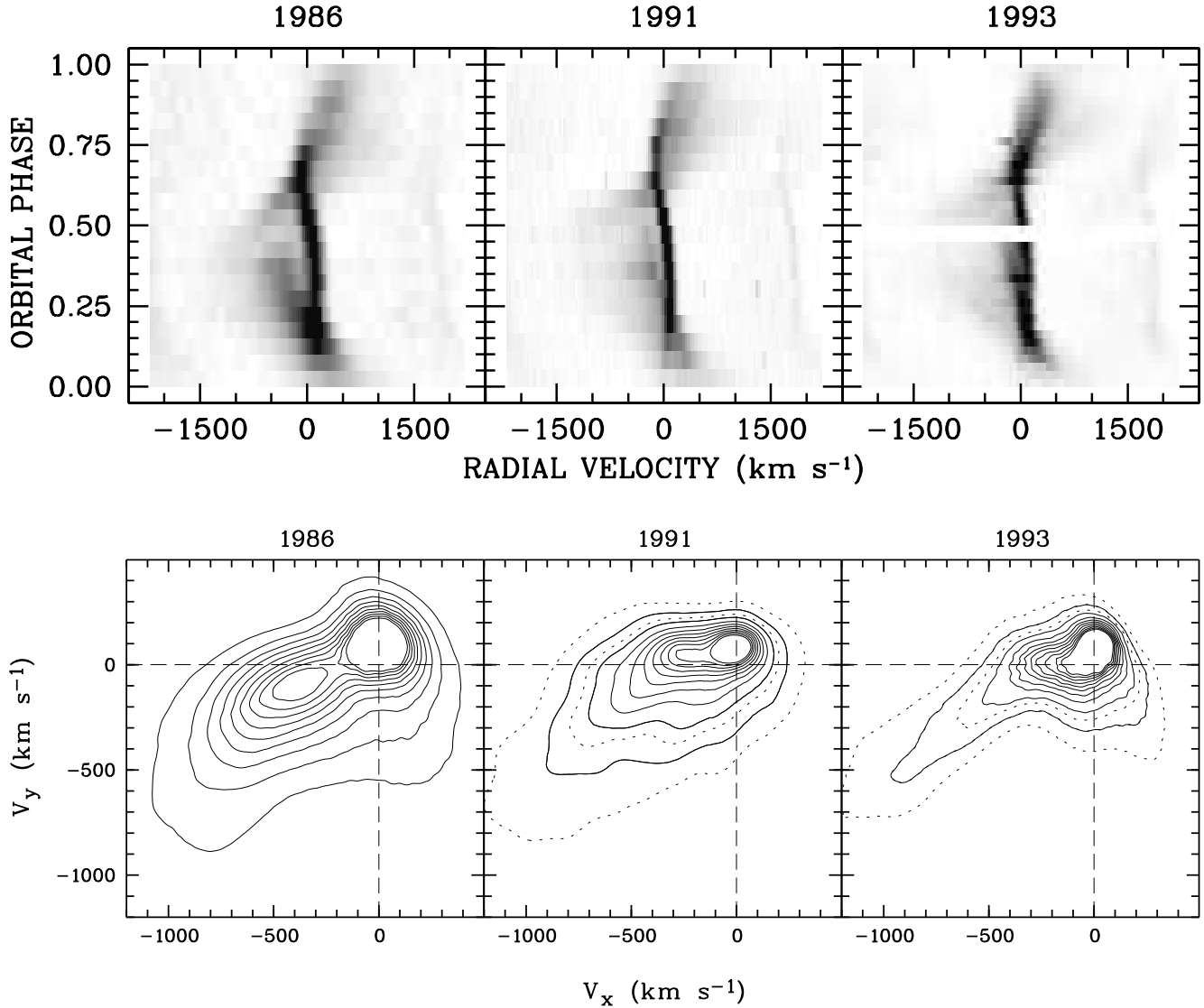}
\caption{Trailed spectrograms (top) and MEM-reconstructed Doppler maps 
(bottom) of the \hel2\ emission lines in 
1986, 1991 and 1993. All spectra were continuum-subtracted and 
phase-averaged according to the ephemeris given in Eq.~1. The dynamic range 
of the grey-scale representations is between 0 and
$150 \times 10^{-16}$\,erg\,cm$^{-2}$\,s$^{-1}$\,\AA$^{-1}$, 
the same in each panel.
Isocontour lines drawn with solid lines 
represent emission at 10\%, 20\% \dots 90\%
of maximum emission originating from the accretion 
stream. For the 1991, and 1993 tomograms contours
at the 5\% and 15\% level are shown in addition drawn with dashed lines.}
\label{f:3mems} 
\end{figure*}

Also the trailed spectrograms and the Doppler maps are markedly 
variable between the three epochs. 
We show in Fig.~\ref{f:3mems} the trails and maps of \heII.
The data of the other emission lines look 
very similar as far as the velocity pattern (though not the brightness pattern)
is concerned. 
Looking first at the trailed spectra, the narrow emission line
(NEL) is the most obvious and the most 
pronounced feature in all three observations. The
underlying emission from the stream
is brighter close to the NEL-emission in 1991 and 1993 than in 1986.
Hence, during the two recent observations the stream was brighter in the 
vicinity of the inner Lagrangian point $L_1$ at rather lower velocities than
in 1986, when stream emission was bright at higher velocities.

This is nicely illustrated too in the Doppler maps which are
shown as contour plots below the trailed spectrograms in Fig.~\ref{f:3mems}.
Contour levels were chosen appropriately 
to emphasize stream emission (the inner contour corresponds to the 
maximum intensity level from the stream).
In 1991 and 1993, there is some bright emission 
originating approximately at $L_1$
and extending down to velocities $v_x \sim -500$\,\kmps\ at constant $v_y$.
From there  a tail stretches  down to the lower left region of the
maps which can be recognized at velocities as high as 
$(v_x,v_y) = (-1200, -700)$~\kmps. 
Within the  resolution of our data,
the orientation of these high-velocity 
tails is the same in all maps.

\begin{figure*}
\psfig{figure=3trails_ps,width=\textwidth,bbllx=50pt,bblly=280pt,bburx=583pt,bbury=500pt,clip=}
\end{figure*}
 
The initial ridge of emission between $L_1$ and 
$(v_x,v_y) \sim (-400, 50)$\,\kmps\ 
can be tentatively identified with emission 
originating from the ballistic stream, as
seen as prominent features in HU~Aqr and UZ~For (Schwope et al.~1997, 1999). 
There are, however, some problems with this interpretation,
since this ridge shows a 
displacement by $\Delta v_y \simeq 45$\,\kmps\ between 1991 and 1993. 
If it were a purely ballistic stream, it would be fixed in the binary 
and always
appear at the same $v_y$--velocity. In the following, we will use the 
term `ballistic' stream for the bright ridge of emission connected to 
the secondary star seen in the 1991 and 1993 tomograms, although it might 
not appear at a ballistic velocity at all.

\begin{figure}
\psfig{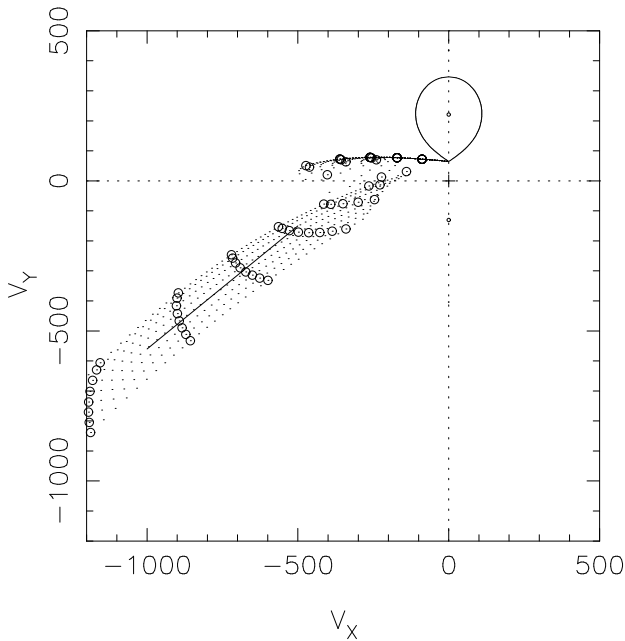}
\caption{Model of the accretion stream in Doppler coordinates.
The adopted 
orientation of the dipole $\vec{\mu}$ is given by its azimuth $\chi_\mu = 
-50\degr$ and its colatitude $\delta_\mu = 23\degr$, the same as in 
Fig.~\ref{f:viewwd}. With this geometry the ballistic part of the stream 
is truncated at $v_x \simeq -500$\,\kmps. The magnetically controlled
part of the flow appears in the lower left quadrant. The straight line 
at the centre of the bundle of trajectories indicates the 
location of the observed stream.
}
\label{f:map_mod} 
\end{figure}

We determined the location of the ballistic stream 
by fitting a Gaussian curve to the Doppler maps
at $v_x = -250$\,\kmps. Maximum emission 
ocurred at $v_y \simeq 45$\,\kmps\ in 1991 and at
$v_y \simeq 0$\,\kmps\ in 1993.
For a first-order estimate we set the $v_y$-velocities 
at $L_1$ and at $v_x = -250$\,\kmps\ to be the same. 
For our best guess of $Q$ and $i$, the 
projected velocity $v_y$ at $L_1$ is 
$75 - 80$\,\kmps\ (see above). Only for the lower limit value
$Q = 1.5$, $i = 70\degr$, is the projected $v_y = 51$\,\kmps\ as low 
as observed in 1991. 
Hence, on both occasions (1991 likely, 1993 definitely), 
the observed ballistic stream
appears at smaller velocities $v_y$ than predicted by a single-particle 
trajectory for the likely parameter combination of $i$ and $Q$.

Emission which could be apparently 
associated to the ballistic stream 
was almost completely absent in 1986. At that 
epoch a bright spot of emission centred on
$(v_x,v_y) \simeq (-380, -100)$\,\kmps\ 
emerged instead. 

As a first step to gain a quantitative understanding of the 
location and extent of the different parts of the stream in the 
magnetosphere, we have computed models for the accretion 
stream, representing it by a single-particle trajectory under gravitational
and centrifugal influence. 
We accounted for the pull of the magnetic field on the 
partly ionized stream by a 
magnetic drag force decreasing exponentially as a function of radius.
This additional drag redirects the trajectory in a small region in 
physical space. It initially follows a ballistic path and then follows
a dipolar field line. The velocity component along the field is conserved 
in our computation when the particle latches onto the field line.
Depending on the orientation 
of the magnetic field in the threading region, the re-direction of the stream
may result in a rather large step in velocity space $(v_x, v_y)$ (see Schwope
et al.~1997 for further details).

A model which accounts for most of the observed features of the 1991 and 1993
tomograms of \hel2 is shown in Fig.~\ref{f:map_mod}. We simulated several 
different trajectories coupling at different radii by varying the 
magnetic field strength, which appears as a dimensionless tuning 
parameter in our calculations. The trajectories shown in Fig.~\ref{f:map_mod}
in velocity space are the same as those shown in Fig.~\ref{f:viewwd} in real 
space. The smaller the assumed value of the field strength, the longer 
the particle remains on the ballistic trajectory (coupling at larger 
azimuth) and the higher the velocity at the threading point becomes. 
Observationally, the ballistic part of the accretion stream can be followed 
down to $v_x \simeq  -500$\,\kmps. It is perhaps a bit
more elongated in 1991 than in 1993, but this is difficult to state
exactly, because
the Doppler map of 1991 has a poorer velocity resolution.
A velocity of 
$v_x \simeq  -500$\,\kmps\ on the ballistic stream corresponds 
to an azimuth of the coupling region of about $-25\degr$. 
Coupling may -- in principle -- occur all along the ballistic stream 
between the $L_1$ point and that azimuth.
The orientation of the axis of the magnetic field was chosen in accord with 
polarimetry (Sect.~\ref{s:polari}). With the chosen field geometry the 
magnetically funnelled part of the trajectories stretches along a more
or less straight line in the lower left quadrant of the velocity plane, 
almost exactly along the observed high-velocity ridge in the Doppler maps
(indicated by a straight line in Fig.~\ref{f:map_mod}). 

This simple model cannot account for the observed displacement towards 
smaller $v_y$-velocities of the `ballistic' stream with respect to the 
single-particle trajectory. One possible explanation for this displacement
is that a proper ballistic stream (not influenced by the magnetic 
field) exists but that the observed emission is dominated by matter 
with slightly changed velocity (matter in the process of getting coupled).
Another possibility has recently been worked out by Sohl \& Wynn (1999)
who treated the accretion flow as a number of diamagnetic blobs which 
independently interact with the magnetosphere of the primary. This model
predicts blob trajectories differing from single-particle trajectories
and might be applicable to \qq.

The apparent absence
of a ballistic stream in 1986 and the 
occurrence of the bright spot at $(v_x,v_y) \simeq (-380, -100)$\,\kmps\ 
remains unexplained so far.
There might be two explanations for the missing ballistic stream:
either matter couples on the field directly at $L_1$ 
or the stream was outshone by other emission
components and therefore not recognizable at the phase and spectral
resolution of our observations. 
We investigated the first possibility by tuning the magnetic field 
strength in our models to high values (which mimics low accretion rates), 
so that no ballistic stream was formed. 
We then used different orientations of the dipolar 
axis in order to check whether the trajectory runs through the observed 
bright spot down to the lower left quadrant in the $(v_x,v_y)$ plane. 
Azimuth and latitude of the 
magnetic axis were varied over large ranges but no solution was found.
We conclude that the resolution of the 1986 data and the brightness of the
other components prevented us from detecting emission from 
the ballistic stream.

\subsubsection{A simple model for the emission line light curves}

Support for the view that the bulk of matter does not couple at $L_1$ 
but somewhere downstream comes from some numerical experiments which were
performed to reach a basic understanding of the emission line light curves.
A single-particle trajectory between the $L_1$ point, the coupling region 
and the white dwarf 
was divided in 100 parts of equal length and each 
element was assigned an intensity value. We allowed for two different 
brightness values, depending on whether the irradiated or the non-irradiated 
side of the stream was seen. Emission from an element was assumed
to be optically thick, hence the observed brightness of an element at 
a given phase was the intrinsic brightness times a foreshortening factor.
The brightness $b$ of the vertical stream (along the dipolar field line)
was varied as a function of radius with $b \propto r^{1.5}$ starting with 
some initial value in the threading region. This radial dependence accounts
for the shrinkage of the effectively emitting area due to the 
converging field lines. 
Light curves were computed by coadding over 
all visible elements at a given phase taking the full Roche geometry 
into account so that eclipses of parts of the stream are recognized 
if the inclination is high enough.

\begin{figure}
\psfig{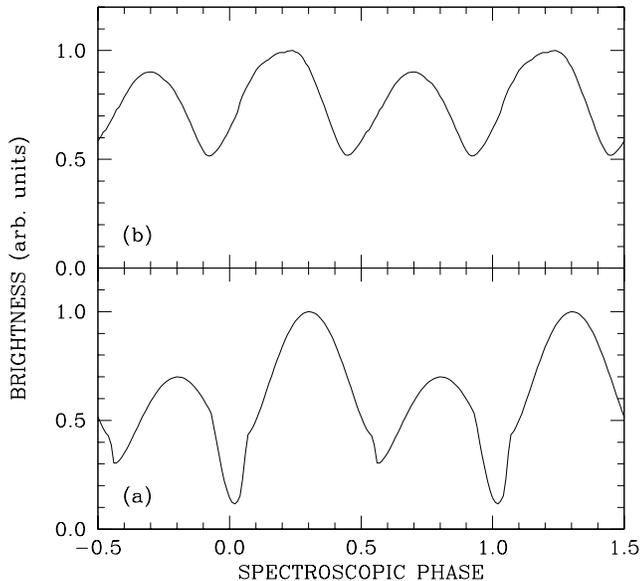}
\caption{Model of the stream emission as function of orbital phase
for the ballistic part in (a) and the magnetically funnelled part in (b).
}
\label{f:lcstr_mod} 
\end{figure}

In Fig.~\ref{f:lcstr_mod} we show the resulting model light curves 
for the ballistic part of the stream and the magnetically funnelled part. 
The central trajectory of those shown in Figs.~\ref{f:map_mod} and 
\ref{f:viewwd} was used and a contrast of $1:0.7$ between 
the irradiated and non-irradiated sides of the stream was adopted. 
The light curves were 
computed for an assumed orbital inclination of 65\degr. Both light
curves were normalized to maximum emission. Emission from the ballistic 
stream shows emission maxima at $\fisp \simeq 0.3$ and 0.8 which is 
simply an expression of the deflection angle of $\sim$18\degr\ of the 
ballistic stream with respect to the line connecting both stars. 
The primary minimum 
at $\fisp \simeq 0.02$ is caused by the combined effect of foreshortening 
and a partial eclipse of the stream by the secondary star. 

The model light curve from the magnetically funnelled part of the stream
shown in Fig.~\ref{f:lcstr_mod}b is affected only by foreshortening. 
This causes the maxima of emission to appear at 
spectroscopic phases 0.2 and 0.7 and the minima at phases 
0.45 and 0.95. These phases are well in agreement with the observed 
maxima and minima of the emission lines of Fig.~\ref{f:3lcs}. 
These phasings  and the absence of pronounced 
stream eclipses by the secondary star are suggestive of stream 
emission mainly originating from the magnetic part of the stream 
(in 1991 and 1993) or the threading region (1986). The visual
impression of the Doppler maps (1991, 1993) is in apparent contradiction 
to this statement, since they show emission associated with the 
ballistic stream as a prominent feature. 
We regard the contradiction as apparent only for 
two reasons. Firstly, as  discussed  above, at least part of 
the emission associated 
with the ballistic stream originates from matter which is
coupled to the field and thus is expected to produce a light curve
resembling that in Fig.~\ref{f:lcstr_mod}. Secondly, 
the surface brightness distribution of a Doppler map does not
reflect the brightness distribution in real space. In particular, emission 
at high velocities is spread over a large region in a $(v_x,v_y)$ diagram.
The summed flux of the high surface-brightness pixels in the region 
defined by $(v_x,v_y) = (-500,-40)$\,\kmps\ (lower left corner) and 
$(120,200)$\,\kmps\ (upper right corner), i.e.~including the NEL from the 
secondary star, contributes only to about 40\% (1991) and 50\% (1993) of the 
total intensity in the maps. Hence, low surface-brightness emission from
higher velocities is dominating in the stream.

The Doppler maps indicate that we can best differentiate between emission 
from the ballistic and the magnetic stream at phases 
0.25 and 0.75, i.e.~in the direction of the $v_x$-axis. Projection in these
directions produces
a relatively sharp spectral feature of emission associated with the ballistic 
stream (although not as sharp as the NEL from the secondary)
whereas emission from the magnetic stream will appear as a
rather broad feature. Fitting the observed spectra (1991 and 1993) 
at these indicated phases with triple Gaussians (NEL plus medium width 
plus broad component) reveals that the medium width and broad base 
components carry almost the same flux at these phases on both occasions.
If we then take into account, that even the medium width component 
is fed by emission from matter which has already latched onto the field
it seems plausible that most of the observed line emission comes 
from the magnetically funnelled stream and/or matter which is in the 
process of becoming funnelled, and that emission 
from the ballistic stream plays a minor 
role in the formation of the emission line light curves.
Hence, Doppler maps and emission line light curves tell us the same story.

\section{Conclusions and outlook}
Using medium- and high-spectral resolution observations with full phase
coverage of the long-period AM Herculis binary \qq\ 
obtained on three occasions in 1986, 1991 and 1993 
and complementary polarimetry obtained between 1985 and 1988
we studied the accretion geometry in the inner and outer magnetosphere.
First of all, we derived a reliable spectroscopic ephemeris
for inferior conjunction of the secondary star by combining 
\na1 photospheric absorption lines with narrow emission lines 
of reprocessed emission from the X-ray irradiated hemisphere,
HJD($T_0$) $= 244\,8446.4710(5) + E \times 0\fd15452011(11)$.
The  arrival times of linear polarization pulses 
show a large scatter in an $(O-C)$ diagram with respect to the 
new ephemeris and thus are disqualified as useful tracers of the 
orbital motion for the case of \qq. 

Analysis of the linear and circular polarization curves favour an
orientation of the (dipolar) magnetic axis of $\delta_\mu =23\degr$
and $\chi_\mu = -50\degr$, and an orbital inclination between $50\degr$ and 
$70\degr$. The variable shape of the polarization curves, in particular 
the phasing of circular polarization dips, 
suggests that the azimuth of the coupling (threading) region undergoes
significant migrations in azimuth. 

The suggested orientation of the magnetic dipole and of the coupling region
is in good agreement with the results inferred from Doppler tomography 
of bright emission lines, in particular those of \heII. These show 
(in 1991 and 1993) three distinct emission features: (1) the irradiated
hemisphere of the secondary star, (2) emission which can be associated 
with the ballistic stream, and (3) emission from the magnetically funnelled
stream. Emission, which apparently arises from the ballistic stream,
appears at different places in the Doppler tomograms of 1991 and
1993. This suggests that it is formed by matter which is in the process
of becoming threaded, in complete agreement with the emission line 
light curves, which show the major contribution from the magnetic part 
of the stream. At one occasion (1986) the 
Doppler maps do not reveal any emission that could be associated with
the ballistic stream.

We explored whether it would be possible to use the narrow emission lines 
from the secondary star in combination with photospheric absorption 
lines to constrain the mass ratio and the orbital inclination.
This was not possible in \qq\ because the 
radial velocity amplitudes of different atomic species were found to be 
widely different. 
Lines from high ionization species such as \heII\ 
originate more closely to the $L_1$ than a reprocessing model predicts.

The Doppler tomograms presented in this paper suggest that threading of the 
accretion stream (or parts of it) starts very soon or perhaps immediately 
after leaving the secondary star. The stream seems to be completely disrupted 
at an azimuth of about $-30\degr$.
Future observations with higher spectral and temporal resolution of 
the \na1-lines (1) will
allow the mass ratio to be determined 
with an accuracy of better than 10\% by the straightforward application 
of Doppler tomography and (2) will
reveal substructure in the Doppler maps due to e.g.~star spots.

\section*{Acknowledgements}
We thank Rick Hessman for carefully reading an early version of the 
manuscript and the anonymous referee for helpful criticism. 

This work was supported by the DLR under grant 50 OR 9403 5
and the Deutsche Forschungsgemeinschaft under grant Be 470/12-2.

\label{lastpage}


\begin{thebibliography}{}
\bibitem[]{}
	Beardmore, A.P., Ramsay, G., Osborne, J.P., Mason, K.O., Nousek, J.A., 
	Baluta, C., 1995, MNRAS 273, 742 
\bibitem[]{} Beuermann, K., Thomas, H.-C., 1990, A\&A 230, 326
\bibitem[]{}
	Brainerd J.J., Lamb, D.Q., 1985, 
     	In: Lamb D.Q., Patterson, J. (eds.) Proc. 7$^{th}$ North American   
     	Workshop on CV's and LMXBs, Reidel, Dordrecht, p.~247
\bibitem[]{}
	Caillault, J.--P., Patterson, J., 1990, AJ 100, 825
\bibitem[]{}
	Catal\'an, M.S., Schwope, A.D., Smith, R.C., 1999, 
	MNRAS, in press (Paper I)
\bibitem[]{}
	Catal\'an, M.S., Davey, S., Smith, R.C., Jones, D.H.P., 1996, 
	in {\it Cataclysmic Variables and Related Objects}, Proc.~IAU
	Coll.~158, Kluwer, Dordrecht, p.~227
\bibitem[]{}Cropper, M., 1989, MNRAS 236, 935
\bibitem[]{}Cropper, M., 1998, MNRAS 295, 353
\bibitem[]{}
	Hauschildt, P.H., Baron, E., Allard, F., 1997, ApJ 483, 390
\bibitem[]{} 
	Liebert, J., Stockman, H.S., 1985, in {\it Proc.~7$^{th}$ 
	North American Workshop on CVs and LMXBs}, 
	eds. D.Q.~Lamb and J.~Patterson, Reidel, Dordrecht, p.~151
\bibitem[]{} 
	Marsh, T.R., Horne, K., 1988, MNRAS 235, 269
\bibitem[]{}
	Meggitt, S.M.A., Wickramasinghe, D.T, 1982, MNRAS 198, 71
\bibitem[]{}
	McCarthy, P., Bowyer, S., Clarke, J.T., 1986, ApJ 311, 873
\bibitem[]{}
	Mukai, K., Charles, P.A., 1986, MNRAS 222, 1P
\bibitem[]{}
	Mukai, K., Charles, P.A., 1987, MNRAS 226, 209
\bibitem[]{}
	Mukai, K., Bonnet-Bidaud, J.-M., Charles, P.A., et al., 1986, 
	MNRAS 221, 839
\bibitem[]{} 
	Neece, G.D., 1984, ApJ 277, 738
\bibitem[]{}
	Nousek, J.A., Takalo, L.O., Schmidt, G.D., et al., 1984, ApJ 277, 682
\bibitem[]{}
	Nugent, J.J., 1983, ApJS 51, 1
\bibitem[]{}
	Osborne, J.P., Beuermann, K., Charles, P.A., Maraschi, L., Mukai, K., 
	Treves, A., 1987, ApJ 315, L123
\bibitem[]{}
	Osborne, J.P., Bonnet-Bidaud, J.-M., Bowyer, S., et al., 1986, 
	MNRAS 221, 823
\bibitem[]{}
	Potter, S.B., Hakala, P.J., Cropper, M., 1998, MNRAS 297, 1261
\bibitem[]{}
	Proetel, K., 1978, PhD thesis, Ruprecht-Karls-Universit\"{a}t, 
	Heidelberg 
\bibitem[]{} 
	Schwope, A.D., Mantel, K.-H., Horne, K., 1997, A\&A 319, 894
\bibitem[]{} 
	Schwope, A.D., Beuermann, K., Jordan, S., Thomas, H.-C., 1993,
        A\&A 278, 498
\bibitem[]{} 
	Schwope, A.D., Beuermann, K., Buckley D.A.H., et al., 1998,
	ASP Conf. Ser.~137, pp.~45--59
\bibitem[]{} 
	Schwope, A.D., Schwarz R., Staude A., Heerlein C., Horne K.,
	Steeghs D., 1999, ASP Conf.~Ser.~157, 71
\bibitem[]{}
	Sohl, K., Wynn, G., 1999, ASP Conf.~Ser.~157, 1999
\bibitem[]{}
	Spruit, H.C., 1998, astro-ph/9806141
\bibitem[]{}
	Stahl, O., Buzzoni, B., Kraus, G., Schwarz, H., Metz, K., 
	Roth, M., 1986,
	The Messenger 46, 23
\end{thebibliography}
\end{document}